\documentclass[12pt]{iopart}

\usepackage{iopams}  
\usepackage{graphicx}
\usepackage[breaklinks=true,colorlinks=true,linkcolor=blue,urlcolor=blue,citecolor=blue]{hyperref}
\usepackage{ulem}
\expandafter\let\csname equation*\endcsname\relax
\expandafter\let\csname endequation*\endcsname\relax
\usepackage{amsmath}
\usepackage{float}
\usepackage{enumitem}
\usepackage{subcaption}
\usepackage{setspace}
\usepackage{cite}
\usepackage{hhline}
\usepackage{blindtext}
\usepackage{subfiles} 
\usepackage{comment}
\newcommand{\Exp}[1]{\big\langle #1 \big\rangle}
\newcommand{\appropto}{\mathrel{\vcenter{
  \offinterlineskip\halign{\hfil$##$\cr
    \propto\cr\noalign{\kern2pt}\sim\cr\noalign{\kern-2pt}}}}}

\begin{document}

\title[\footnotesize An Experiment for observing quantum gravity phenomena using twin table-top 3D interferometers]{An experiment for observing quantum gravity phenomena using twin table-top 3D interferometers}

\author{Sander~M.~Vermeulen, Lorenzo~Aiello, Aldo~Ejlli, William~L.~Griffiths, Alasdair~L.~James, Katherine~L.~Dooley, and Hartmut~Grote}
\address{Gravity Exploration Institute, Cardiff University, Cardiff CF24 3AA, United Kingdom}
\ead{vermeulensm@cardiff.ac.uk}

\graphicspath{{Figures/}{../Figures/}}



\begin{abstract}
Theories of quantum gravity based on the holographic principle predict the existence of quantum fluctuations of distance measurements that accumulate and exhibit correlations over macroscopic distances. This paper models an expected signal due to this phenomenology, and details the design and estimated sensitivity of co-located twin table-top 3D interferometers being built to measure or constrain it. The experiment is estimated to be sensitive to displacements $\sim10^{-19}\,\rm{m}/\sqrt{\rm{Hz}}$ in a frequency band between 1 and 250 MHz, surpassing previous experiments and enabling the possible observation of quantum gravity phenomena. The experiment will also be sensitive to MHz gravitational waves and various dark matter candidates.
\end{abstract}



\section{Introduction}
A physical description of nature including gravity at the Planck scale has not yet been found. Two otherwise successful theories, general relativity and quantum mechanics, provide descriptions that are irreconcilable at these small length and high energy scales \cite{deser_nonrenormalizability_1974,han_scale_2005,burgess_quantum_2004}. Theoretical efforts are ongoing to quantise gravity and address the incompatibilities, and include theories such as string theory and loop quantum gravity \cite{ashtekar_introduction_2013,kiritsis_introduction_1998,hossenfelder_minimal_2013}. These theories usually predict that new physics appears only at scales on the order of the square of the Planck length $l_{\rm{P}}$, making them untestable in current experiments \cite{zurek_does_2019}. 

However, it has been widely established that a general property of theories of quantised space-time is that measurements of distance are subject to an uncertainty principle beyond that of Heisenberg; i.e. space-time exhibits quantum fluctuations. Repeated measurements of a distance will therefore show fluctuations with an irreducible variance \cite{hawking_spacetime_1978,amelino-camelia_phenomenological_2001,amelino-camelia_quantum_2013,ng_measuring_2000,jack_ng_limit_1994,hogan_measurement_2008}. Moreover, the magnitude of these fluctuations of distance measurements $\delta L$ are theorised to scale with the distance $L$ that is measured, i.e. 
\begin{equation}\label{Eq:scaling}
    \delta L \propto (l_{\rm{P}})^{\alpha} (L)^{1-\alpha},
\end{equation}
where $\alpha$ is a constant specific to each model of quantised space-time \cite{amelino-camelia_gravity-wave_1999,ng_selected_2003,christiansen_limits_2011,kwon_interferometric_2016}. 

Additionally, the holographic principle \cite{t_hooft_dimensional_2009,susskind_world_1995,bousso_holographic_2002} and the covariant entropy bound \cite{bousso_covariant_1999} that follows from it imply that these distance fluctuations are correlated in a given volume of space-time. Furthermore, work by Verlinde \& Zurek \cite{verlinde_observational_2019,verlinde_spacetime_2019} and 't~Hooft \cite{t_hooft_quantum_2016,t_hooft_discreteness_2018} suggests that these correlations might extend over macroscopic distances transverse to light-sheets (or equivalently, along boundaries of causal diamonds \cite{banks_holographic_2011}).

These theoretical approaches evaluate quantum fluctuations and their correlations at horizons, and by identifying boundaries of causal diamonds as horizons (specifically Rindler horizons), the transverse correlations of quantum space-time fluctuations may be described.  In particular, Verlinde \& Zurek postulate that energy fluctuations prescribed by the thermodynamic properties of horizons might give rise to metric fluctuations at the horizon with transverse correlations through a Newtonian potential \cite{verlinde_observational_2019}. 't~Hooft proposes that black holes can obey unitarity if the quantum fluctuations at the horizon (e.g. Hawking radiation) are antipodally entangled \cite{t_hooft_black_2016}. These theories provide concrete and almost identical predictions for the angular two-point correlation function of the fluctuations as an expansion in spherical harmonics \cite{verlinde_observational_2019,hogan_nonlocal_2019,t_hooft_quantum_2016}. The decomposition of the correlations into spherical harmonics $Y_l^m$ derived in this way has most of its power in low $l$ modes, which motivates the prediction that the transverse correlations extend over macroscopic angular separations, as mentioned above. Moreover, it has been suggested that the angular power spectrum of the temperature fluctuations in the CMB is a manifestation of this fundamental decomposition in spherical harmonics of quantum fluctuations on an inflationary horizon \cite{hogan_pattern_2019}. 

Importantly, macroscopic transverse correlations imply that fluctuations are coherent over the typical diameter of a laser beam or telescope aperture. If this is the case, astrophysical constraints set on quantum space-time fluctuations by evaluating the blurring or degrading of images of distant objects \cite{perlman_new_2015,christiansen_limits_2011} might not apply. 

Given that quantum space-time fluctuations of distance scale with and exhibit correlations over macroscopic distances, laser interferometers are uniquely sensitive to them. The most stringent constraints on these fluctuations are therefore set by existing interferometry experiments. The design of the gravitational wave interferometers used by the LIGO, VIRGO, and KAGRA collaborations \cite{pitkin_gravitational_2011} reduces their potential sensitivity to quantum space-time fluctuations. This is because they employ Fabry-P\'erot cavities in the arms (or have folded arms, as in GEO\,600), which means individual photons traverse the same distance multiple times. In addition, the output of these instruments is sampled at a frequency that is lower than the light crossing frequency. This causes the random but detectable signal from fluctuations accumulated in a single light crossing to be averaged with the signal of subsequent crossings, negating the effect \cite{kwon_interferometric_2016}.    

An interferometry experiment designed to detect quantum space-time fluctuations is the Fermilab Holometer, which consists of two identical co-located and power-recycled 40 m Michelson interferometers \cite{chou_holometer_2017}. This experiment constrained the magnitude of possible quantum space-time fluctuations in the cross-spectrum of the interferometers to a strain power spectral density less than the Planck time ($t_{\rm{P}}$) \cite{chou_interferometric_2017}.    

However, given the theoretical evidence that quantum space-time fluctuations are different for distances transverse to and along light-sheets and exhibit certain angular correlations (see above), the geometry of the Holometer might greatly reduce its sensitivity to these phenomena, as the interferometer arms are orthogonal to each other and employ only light paths along the light-sheet. 

In general, the prediction that quantum space-time fluctuations are described by a non-constant angular two-point correlation function implies that the signal due to these fluctuations in interferometers depends on the angle between the arms, as an interferometer produces a signal only for anti-correlated changes in the length of its arms. This warrants a reconfigurable interferometer design in which the angle between the arms can be varied between measurements. Moreover, the general prediction that the fluctuations are correlated between different points within the same causally connected volume of space-time implies that separate measurements within that volume are correlated, which means co-located instruments will exhibit cross-correlations in their signals.     

This paper details a new experiment that will be set up at Cardiff University which will consist of twin table-top power-recycled interferometers with reconfigurable three-dimensional arms that define light paths with components along and transverse to boundaries of causal diamonds. The interferometers will be cross-correlated on timescales much shorter than the light crossing time, and can ultimately be sensitive to displacements $\sim10^{-19}\,\rm{m}/\sqrt{\rm{Hz}}$ at high frequencies ($\sim$~MHz), surpassing previous interferometry experiments. The instrument can thus be used to make observations with unprecedented sensitivity to confirm or constrain various theories of quantised space-time. This experiment will also be sensitive to high-frequency gravitational waves, which may be emitted by primordial black hole mergers and other potential components of a stochastic gravitational wave background \cite{chou_mhz_2017}. In addition, data from the interferometers can also be used to constrain parameters of some dark matter candidates through various coupling mechanisms \cite{grote_novel_2019,hall_laser_2018,brito_gravitational_2017,pierce_searching_2018}.  

This paper is structured as follows. In Sec.~\ref{sec:Model_of_Signal}, a simple model is formulated and then used to derive a frequency-domain signal due to quantum space-time fluctuations in a 3D interferometer. Auxiliary science goals, which include the detection of gravitational waves and dark matter, are discussed in Sec.~\ref{sec:Auxiliary_Science_Goals}. In Sec.~\ref{sec:Detection_Statistic}, the detection statistic that quantifies the sensitivity of the experiment is given. The experimental design and improvements over previous experiments are outlined in Sec.~\ref{sec:Experiment_Design}, and relevant sources of noise and strategies for their mitigation are expanded on in Sec.~\ref{sec:Mitigation_of_Noise}. An estimate of the experiment's sensitivity is provided in Sec.~\ref{sec:Projected_Sensitivity}, and the paper is summarised in Sec.~\ref{sec:Conclusions}.


\section{Model of Signals due to Quantum Space-Time Fluctuations in a 3D Interferometer}\label{sec:Model_of_Signal}
The statistical phenomenological prediction for variations in length measurements due to the quantum nature of gravity as summarised in Eq.~\ref{Eq:scaling} is not readily translated to a quantitative prediction for the signals expected in interferometers. Below, a simple model in which the quantum space-time fluctuations affect photon geodesics through perturbations of the Minkowski metric according to this particular statistical phenomenological prediction is used to derive a frequency-domain signal. 

\begin{figure}[ht]
  \begin{subfigure}[b]{0.51\textwidth}
    \includegraphics[width=\textwidth]{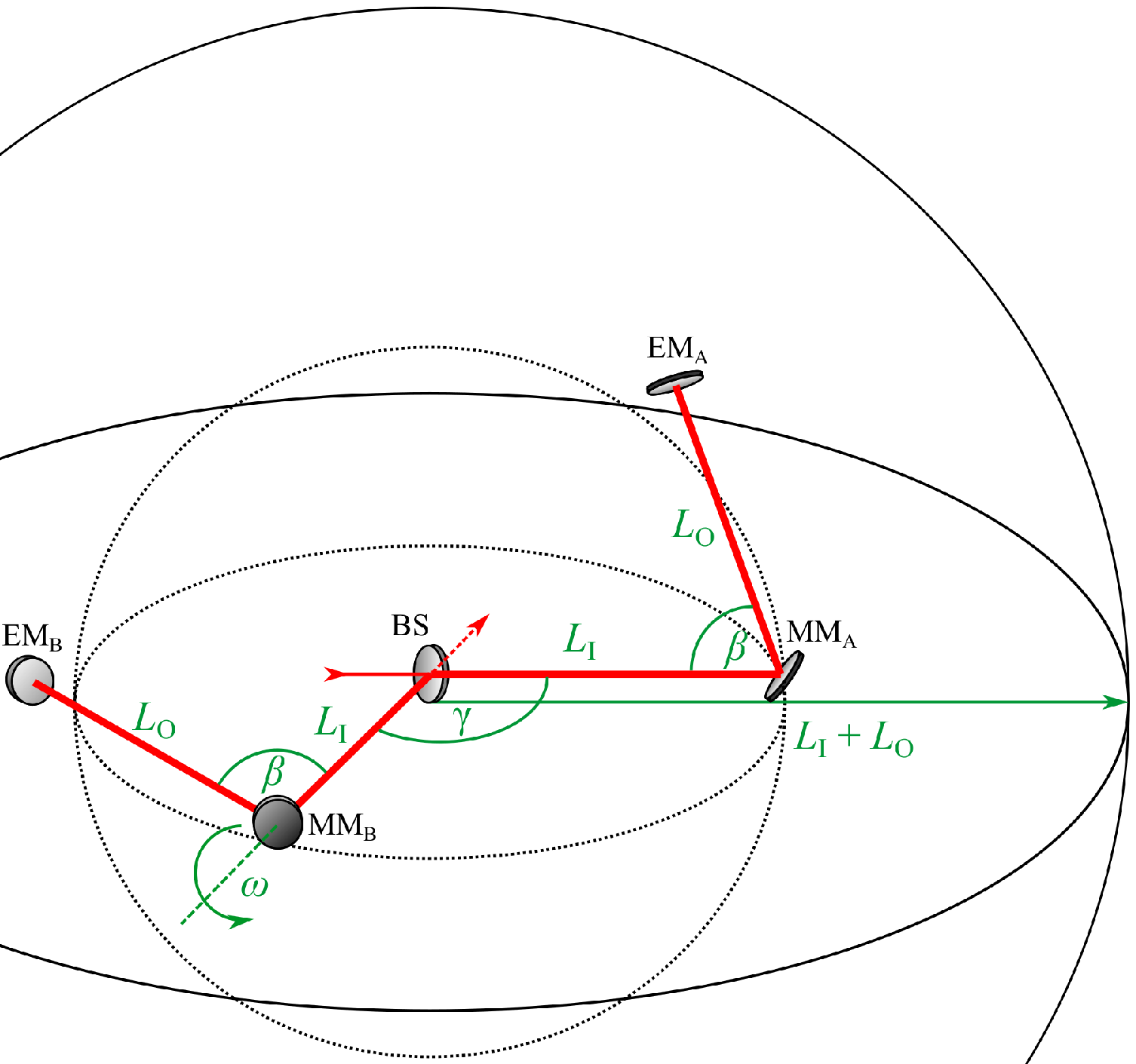}
    \caption{}
    \label{fig:geom_single_space}
  \end{subfigure}
  \hfill
  \begin{subfigure}[b]{0.45\textwidth}
    \includegraphics[width=\textwidth]{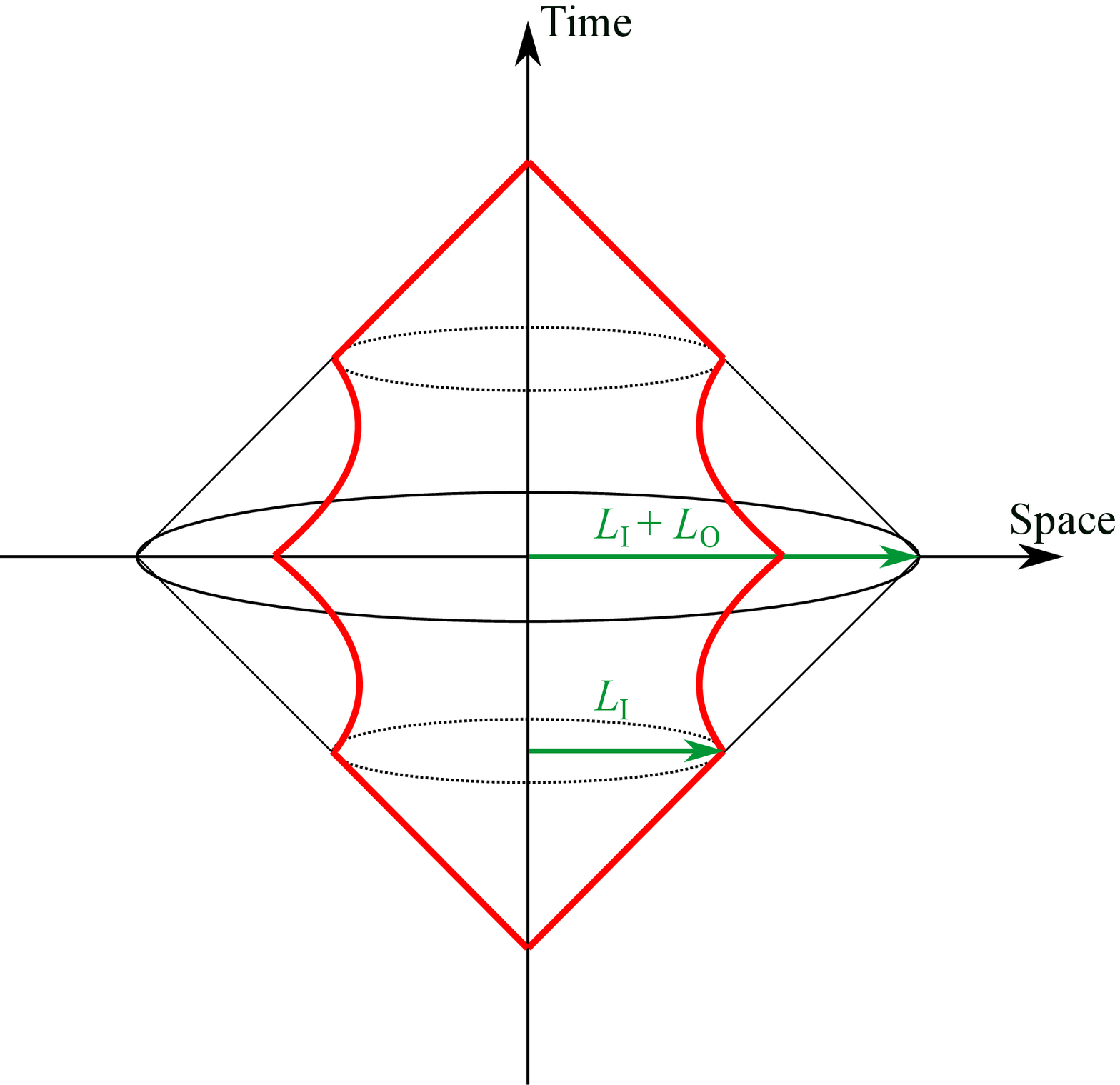}
    \caption{}
  \end{subfigure}
  \caption{\textbf{a)} Schematic of the geometry of a single interferometer in 3 spatial dimensions. Arms (red) consist of an inner radial and outer non-radial segment with lengths $L_{\rm{I}}$ and $L_{\rm{O}}$, respectively. The angle between the inner segments of the two arms is $\gamma$, and the angle between the inner and outer segments in both arms is $\beta$. Light is injected and split into the arms at the beamsplitter (BS), reflected into the outer arm segments by mirrors ($\rm MM_{A,B}$), and returned along the same path by reflection off the end mirrors ($\rm EM_{A,B}$), after which it interferes at the beamsplitter. The outer solid sphere is a 2D space-like boundary of a (3+1)D causal diamond defined by the experiment. This boundary is called the holographic screen of the causal diamond. \textbf{b)}~Space-time diagram of two photons traversing both interferometer arms from beamsplitter to end mirror and back. The photon paths are projected onto the radial spatial direction, and the two transverse spatial directions are suppressed.  The top and bottom vertices of the solid square correspond to the photons leaving and returning to the beamsplitter and define a causal diamond. The boundary of this causal diamond is the space-like circle around the origin, which represents the same surface as the solid sphere in Fig. \ref{fig:geom_single_space}.}
  \label{Fig:geometry}
\end{figure}

\begin{figure}[ht]
  \begin{subfigure}[b]{0.51\textwidth}
    \includegraphics[width=\textwidth]{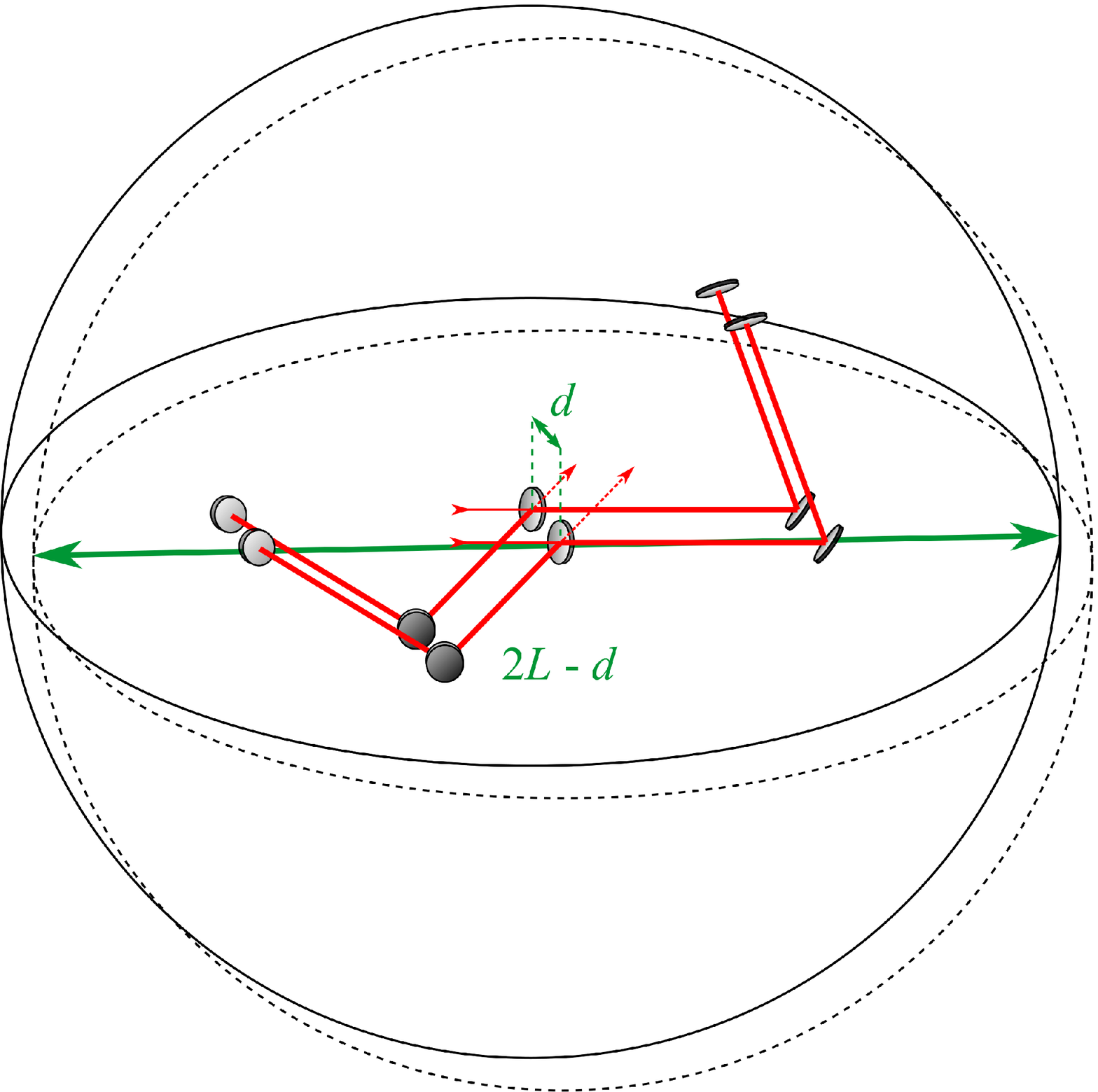}
    \caption{}
    \label{fig:geom_double_space}
  \end{subfigure}
  \hfill
  \begin{subfigure}[b]{0.45\textwidth}
    \includegraphics[width=\textwidth]{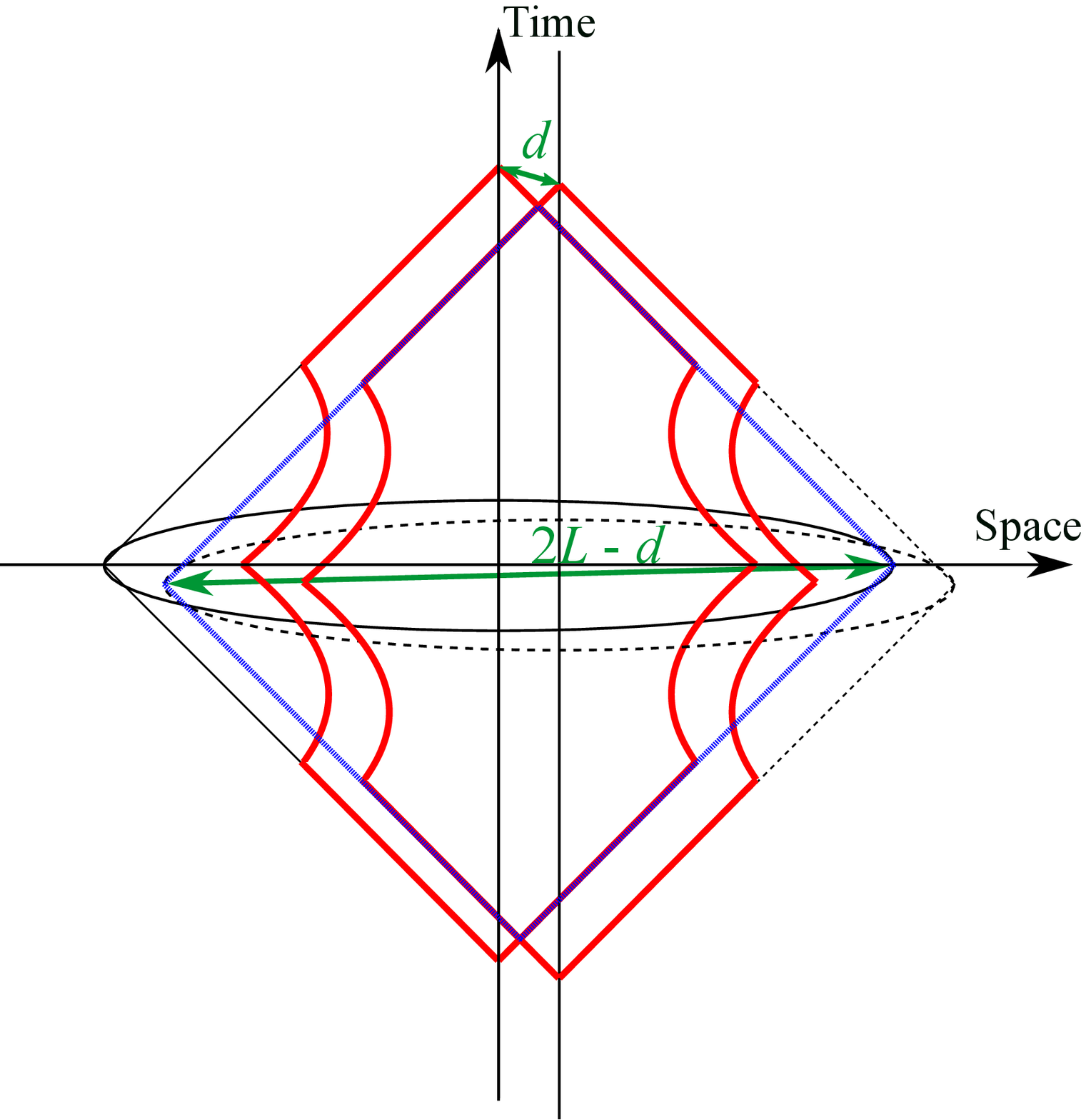}
    \caption{}
  \end{subfigure}
  \caption{\textbf{a)} Schematic of the two identical co-located interferometers with 3D arms of total optical path length $L$. The interferometers are separated by a distance $d$. The spheres are the boundaries of causal diamonds defined by the experiment. \textbf{b)} Space-time diagram of four photons traversing both arms in both interferometers from beamsplitter to end mirror and back. The photon paths are projected onto the radial spatial direction, and the two transverse spatial directions are suppressed. The solid and dashed squares are the causal diamonds defined by the departure and return of the photons from the beamsplitters in either interferometer. The boundaries of these causal diamonds are the solid and dashed space-like circles, which represent the same surfaces as the solid and dashed spheres in Fig. \ref{fig:geom_double_space}. The blue rectangle is the overlap of these two causal diamonds, which is also a causal diamond of size $2L-d$. }
  \label{Fig:geometry2}
\end{figure}

\subsection{Model of Fluctuations of Measured Length}
Consider an interferometer as in Fig.~\ref{Fig:geometry} where the two arms A and B, intersecting at the beamsplitter BS, are at an angle $\gamma$ with respect to each other. The inner, radial, section of both arms has a length $L_{\rm{I}}$. Mirrors $\rm{MM_{A,B}}$ are placed at the end of the first section to reflect the beams of either arm at an angle $\beta$ into an outer, non-radial arm section with length $L_{\mathrm{O}}$. The light in both arms is returned along the same path by end mirrors $\mathrm{EM_{A,B}}$.

Let a photon traverse an arm from beamsplitter to end mirror once. The arm length that is measured at a time $t$ by comparing the phase of such a photon to a stable reference is given by the proper distance between beamsplitter and end mirror, where only the spatial part of the metric is considered \cite{verlinde_observational_2019,rakhmanov_response_2005,schutz_first_1985}: 
\begin{equation} \label{pathlength}
    L(t) = \int_0^L \sqrt{g_{ij}\,\lambda^{i\prime}\lambda^{j\prime}}\,ds,
\end{equation}
where $g_{ij}$ is the metric, $\lambda^i(s)$ is a parameterisation of the photon's path in the parameter $s\in [0, L]$, $\lambda^{i\prime}(s)$ is the tangent vector, roman indices denote spatial components, repeated indices are to be summed over, and $L$ is the full optical path length of the arm at any time. Natural units are used unless specified otherwise. 
 
To model the signal due to quantum space-time fluctuations that might be detected in such an experiment, it is assumed the experiment is done in flat (Minkowski) space-time $\eta_{ij}$, where quantum geometrical fluctuations manifest as small perturbations $h_{ij}$ of the metric;
\begin{equation}\label{perturb}
    g_{ij}= \eta_{ij} + h_{ij}.
\end{equation}
The theoretical origin and magnitude of the quantum space-time fluctuations is not considered here. Using Eq.~\ref{perturb} in a Taylor expansion of Eq.~\ref{pathlength}, the measured length change due to quantum space-time fluctuations to first order can be written as 
\begin{equation}\label{delL}
   \delta L(t)=  \frac{1}{2}\int_0^L h_{ij}\,\sqrt{\lambda^{i\prime}\lambda^{j\prime}}\,ds.
\end{equation}
Note that this expression entails an integral of the fluctuations over a trajectory in the space-time volume of a causal diamond. In a holographic scenario, an integral of such a geodesic would require bulk reconstruction \cite{czech_integral_2015,verlinde_spacetime_2019}. Eq.~\ref{delL} is to be seen as a simplified model of how a length change due to quantum space-time fluctuations might arise.

\subsection{Model of Statistics of Fluctuations}
To be able to model the frequency domain signal that would be obtained if the phase of many photons successively traversing the arm once is measured, some assumptions about the statistics of the fluctuations $h_{ij}(s)$ need to be made. In the case that the fluctuations behave like homogeneous and isotropic white noise with a variance on the order of the Planck length, i.e. \cite{verlinde_observational_2019}
\begin{equation} \label{Eq:whitenoise}
    \Exp{h_{ij}(s_1)h_{lk}(s_2)} =  A^2l_{\mathrm{P}}\delta(s_1-s_2),
\end{equation}
the two-time correlation function of length measurements of any photon path is:
\begin{equation}\label{Eq:WN2timecorr}
\Exp{\delta L(t_1)\delta L(t_2)} = \frac{1}{4}\Bigg \langle \int_0^{L}\int_0^{L}   h_{ij}(s_1) h_{kl}(s_2)  \,ds_1\, ds_2  \Bigg \rangle =  A^2l_{\mathrm{P}}L \Lambda (\tau),
\end{equation}
where angular brackets denote expectation values, $A$ is some constant of order one, $\tau=t_1-t_2 = s_1-s_2$, and $\Lambda (\tau)$ is a triangle function defined as \cite{hogan_statistical_2017-1}
\begin{equation}
    \Lambda (\tau) = \begin{cases}
1-|\tau|/L & \quad 0 < \tau < L \\
0 & \quad \text{otherwise}.
    \end{cases}
\end{equation}

Eq.~\ref{Eq:WN2timecorr} does not allow all the phenomenologies of fluctuations of measured distance proposed in theory to be modelled. Specifically, it can be seen that the case that the space-time fluctuations behave like uncorrelated white noise with Planckian variance as in Eq.~\ref{Eq:whitenoise} leads to a scaling of length fluctuations as in Eq.~\ref{Eq:scaling} with $\alpha=1/2$. Models that predict $\alpha=1/2$ are often categorised as `random walk models' \cite{amelino-camelia_gravity-wave_1999,christiansen_limits_2011}, and this value for $\alpha$ has also been predicted more recently in theoretical work that invokes the holographic principle \cite{verlinde_observational_2019,hogan_interferometers_2012}. Other, older theories inspired by holography predict $\alpha=1/3$ \cite{ng_measuring_2000,karolyhazy_gravitation_1966,singh_quantum_2020}. Finally, more conservative approaches to quantum gravity such as string theory and loop quantum gravity predict no accumulation of fluctuations over distance, which implies $\alpha=1$ \cite{garay_quantum_1995,hossenfelder_minimal_2013,amelino-camelia_quantum_2013}. 

To reproduce these different phenomenological predictions for the scaling of measured length changes summarised in Eq.~\ref{Eq:scaling}, to account for homogeneous but anisotropic quantum space-time fluctuations \cite{hogan_statistical_2017}, and to allow the modelling of different arm geometries, it might be assumed that in general 
\begin{align}\label{Eq:autocorr}
    \Exp{\delta L(t_1)\delta L(t_2)}&=  \frac{1}{4}\Bigg \langle \int_0^{L}\int_0^{L}   h_{ij}(s_1)\sqrt{\lambda^{i\prime}\lambda^{j\prime}} \; h_{kl}(s_2) \sqrt{\lambda^{k\prime}\lambda^{l\prime}}\;  ds_1 ds_2  \Bigg \rangle \\ &= \Exp{\lambda^{m\prime}\lambda^{n\prime}}A_{mn}^2(l_{\rm{P}})^{2\alpha} (L)^{2(1-\alpha)}\Lambda (\tau), \label{Eq:varscaling}
\end{align}
where $A_{mn}\equiv \Exp{|h_{mn}|}$. The expectation value of the magnitude of the length fluctuations, 
\begin{equation}
    \Exp{|\delta L(t)|}\approx \sqrt{\Exp{\delta L(t)\delta L(t)}}  ,
\end{equation}
thus scales according to the phenomenological predictions in Eq.~\ref{Eq:scaling} as desired: 
\begin{equation}\label{Eq:exp_mag_dL}
    \Exp{\left|\delta L(t)\right|}=  \frac{1}{2}\Bigg \langle  \left| \int_0^{L}  h_{ij}\,\sqrt{\lambda^{i\prime}\lambda^{j\prime}}  \,ds  \right|\Bigg \rangle \approx \left(\Exp{\lambda^{i\prime}\lambda^{j\prime}}A_{ij}^2\right)^{\frac{1}{2}}(l_{\rm{P}})^{\alpha} (L)^{1-\alpha}.
\end{equation}
Setting $\alpha=1/2$ and $A_{ij}=A\delta_{ij}$ in this expression corresponds to the case of homogeneous isotropic white noise. Note that for $\alpha\neq 1/2$, Eq.~\ref{Eq:varscaling} implicitly assumes a two-point correlation function for quantum space-time fluctuations different from Eq.~\ref{Eq:whitenoise}. 

Eq.~\ref{Eq:exp_mag_dL} can now be evaluated for an interferometer arm geometry as in Fig.~\ref{Fig:geometry} in a spherical basis ($i \in \{r,\theta,\phi\}$):
\begin{align}
       \Exp{\left|\delta L(t)\right|} &= \frac{1}{2} \Bigg \langle \int_0^{L_\mathrm{I}} \lambda'_{r_\mathrm{I}}  h_{rr} \,ds_\mathrm{I} + \int_0^{L_{\mathrm{O}}} h_{ij}\,\sqrt{\lambda^{i\prime}_{\mathrm{O}} \lambda^{j\prime}_{\mathrm{O}}} \,ds_{\mathrm{O}} \Bigg \rangle \\
       &=A_{rr}(l_{\rm{P}})^\alpha (L_\mathrm{I})^{1-\alpha} + \left(\Exp{\lambda^{i\prime}_{\mathrm{O}}\lambda^{j\prime}_{\mathrm{O}}}A^2_{ij}\right)^{\frac{1}{2}}(l_{\rm{P}})^\alpha (L_\mathrm{O})^{1-\alpha},
\end{align}
where the subscripts $\rm{I,O}$ refer to the inner and outer arm segments respectively. It can be seen from this expression that the inner arm segment is only sensitive to radial fluctuations ($h_{rr}$), which are those along light-sheets (transverse to boundaries of causal diamonds). The outer arm segment is in general sensitive to fluctuations in any direction, including those transverse to light-sheets ($h_{\theta\theta}$ and $h_{\phi\phi}$, along boundaries of causal diamonds).        

\subsection{Frequency Domain Signal}
The frequency domain signal produced by the quantum space-time fluctuations, specifically the power spectral density of the length fluctuations, can be found by taking the Fourier transform of the two-time correlation function in Eq.~\ref{Eq:autocorr}:
\begin{equation}
       P_{xx}(f)= \int_{-\infty}^{\infty} \Exp{\delta L(t_1)\delta L(t_2)}\, \rme^{-2\pi \rmi f \tau} \,d\tau , 
\end{equation}
which gives,
\begin{equation}\label{Eq:PSD}
   P_{xx}(f)= \frac{l_{\rm{P}}^{2\alpha}}{c}\left(   A_{rr}(L_\mathrm{I})^{(\frac{3}{2}-\alpha)} \text{sinc}\left(\frac{\pi L_\mathrm{I} f}{c}\right) + \left(\Exp{\lambda^{i\prime}_{\mathrm{O}}\lambda^{j\prime}_{\mathrm{O}}}A^2_{ij}\right)^{\frac{1}{2}} (L_\mathrm{O})^{(\frac{3}{2}-\alpha)}\text{sinc}\left(\frac{\pi L_\mathrm{O} f}{c}\right) \right)^2,
\end{equation}
in SI units ($\mathrm{m}^2/\mathrm{Hz}$), and the amplitude spectral density is the square root of  $P_{xx}(f)$. When evaluated for $\alpha=1/2$, Eq.~\ref{Eq:PSD} has the same form as previous results for quantised space-time signals in interferometers that consider the case $\alpha=1/2$ \cite{hogan_statistical_2017-1,verlinde_observational_2019,hogan_statistical_2017}. This expression does not take into account the second traversal of the arm from end mirror to beamsplitter, and the interference with light from the second arm. Moreover, this model does not yet incorporate the correlations or entanglement of the fluctuations implied by theory.  

\subsection{Cross-Spectral Density Signal in Co-Located Interferometers}
In theory, the signal described by Eq.~\ref{Eq:PSD} is detectable in a single instrument. However, there is strong theoretical evidence, founded on the holographic principle, that measurements within the same causally connected volume of space-time are correlated. Therefore, the signal due to quantum space-time fluctuations in a single instrument should be coherent to some degree with the signal in an identical interferometer in the same volume of space-time. As dominant sources of noise are expected to have limited coherence between the interferometers, the attainable signal-to-noise ratio for quantum space-time fluctuations is expected to be much greater in the time-integrated cross-spectral density of such co-located instruments than in the auto-spectrum of a single interferometer. In a simple holographic scenario, it may be expected that the strength of the correlations of these measurements is proportional to the overlapping volume of the causal diamonds defined by the respective measurements (see Fig.~\ref{Fig:geometry2}) \cite{banks_holographic_2011,hogan_statistical_2017-1}. In the unrealistic approximation that the interferometers are truly co-located, the causal diamonds defined by the two interferometers overlap fully, and the cross-spectral signal is equal to the auto-spectral signal (Eq.~\ref{Eq:PSD}). The overlap of the individual causal diamond is itself a causal diamond of size $2L-d$, where $d$ is the separation between the interferometers and $2L$ is the size of causal diamonds defined by a single interferometer \cite{gibbons_geometry_2007}. Imperfections of the identicality of the twin interferometers, specifically in the lengths and angles, may also decrease the overlap of the causal diamonds, but these geometrical imperfections can easily be limited such that their effect is much smaller than the effect of the separation between the interferometers. The signal in the cross-spectrum is thus expected to decrease as $\sim -2d/L$ to first order for inter-instrument separations that are small compared to the spatial radii of the individual causal diamonds.


\section{Auxiliary Science Goals}\label{sec:Auxiliary_Science_Goals}
The co-located 3D interferometers described in this paper will have unprecedented sensitivity to high-frequency differential fluctuations in the length of the arms, especially if these changes are coherent across the two instruments. Therefore, in addition to the above, any phenomenon that couples to the apparatus such that it induces coherent length fluctuations at these frequencies can be detected or constrained. This makes the experiment well-suited to probe a wide, previously unexplored, range of physical theories. Phenomena of particular interest and potential, which the experiment may detect or constrain beyond current limits, are gravitational waves and dark matter. 

\subsection{Gravitational Waves}
Gravitational Waves (GWs) are predicted to exist at virtually all frequencies, although to date only GWs emitted by the coalescence of black holes and neutron stars with frequencies ranging roughly between several Hz and one kHz have been detected \cite{ligo_scientific_collaboration_and_virgo_collaboration_gwtc-1_2019}. The co-located 3D interferometers will be uniquely sensitive to high-frequency ($\sim$~MHz, see Sec.~\ref{sec:Projected_Sensitivity})  GWs, as predicted to be emitted by various astrophysical sources.

Specifically, high-frequency GWs are likely present in an unresolved stochastic background \cite{cruise_potential_2012}, as they may be produced by e.g. primordial black holes \cite{wang_constraints_2018,carr_black_1974}, cosmic strings \cite{siemens_gravitational-wave_2007}, anisotropic inflation \cite{ito_mhz_2016}, and other sources in the early universe \cite{bisnovatyi-kogan_very_2004}. The merging of binary primordial black holes that exist to this day may also produce high-frequency GWs \cite{raidal_gravitational_2017,chou_mhz_2017}. In addition, it has been theorised that superradiant instabilities of rotating black holes due to undiscovered low-mass fields (dark matter) can be a source of monochromatic gravitational waves \cite{arvanitaki_detecting_2013,rosa_stimulated_2018}, and the experiment would be senstitive to these GWs for certain parameter ranges. 

While detections are not very likely \cite{martinez_garcia_cors_searching_2020}, it seems worthwhile to explore this frequency range and develop the interferometer technology to unprecedented sensitivity.  

\subsection{Dark Matter}
Various coupling mechanisms between interferometers and a range of microscopic dark matter (DM) candidates, proposed within the framework of quantum field theory as extensions of the standard model of particle physics, have been quantitatively described in the literature. These approaches generally describe undiscovered weakly interacting, low-mass (sub-eV) fields, known as WISPs or VULFs \cite{ringwald_exploring_2012,essig_dark_2013,derevianko_detecting_2018}. One class of theories postulates non-gravitational couplings of light scalar or pseudo-scalar fields to the electromagnetic field and fermions, whereby fundamental constants oscillate \cite{stadnik_can_2015,stadnik_searching_2015,arvanitaki_sound_2016,stadnik_enhanced_2016}. This causes the size of interferometer components to oscillate, producing a signal at a frequency set by the mass of the field \cite{grote_novel_2019,geraci_searching_2019}. Others hypothesise how new vector bosons could exert an oscillatory force on interferometer components, producing a similar oscillatory signal at a frequency determined by the boson's mass \cite{pierce_searching_2018}. In addition, undiscovered low-mass fields could form topological defects, which could produce a transient signal through the interaction between the interferometer and a passing defect \cite{derevianko_hunting_2014}. As mentioned above, low-mass fields, specifically axion-like particles, which are also weakly interacting massive DM candidates, can be indirectly detected from GWs emitted through black hole superradiance.

All these signatures of microscopic DM are expected to have macroscopic spatial coherence, and therefore the co-located interferometers will be uniquely sensitive to such signatures at high frequencies. This will allow the exploration of previously unconstrained areas of the parameter space of these DM candidates \cite{derevianko_detecting_2018}.          

Macroscopic DM candidates that couple with interferometers gravitationally, non-gravitationally, or both, have also been considered in the literature. One category of such scenarios entails detectable local stochastic interactions between `clumpy' DM and interferometer components \cite{adams_direct_2004,hall_laser_2018,tsuchida_dark_2020}. Primordial black holes are also macroscopic DM candidates, and their existence can be probed through GW emission as described above. 
 
\subsection{Entangled Squeezing}
As a further auxiliary topic of interest, it has been proposed that the sensitivity to signals that are correlated between two interferometers may be greatly improved by injecting a single entangled squeezed (or `twin-beam-like') state of light into both interferometers \cite{ruo_berchera_quantum_2013}. This approach theoretically increases the sensitivity in the cross-spectrum of the two interferometers beyond the photon shot noise limit as encountered using two independent squeezed states \cite{berchera_two-mode_2015}. Recently, entangled squeezing was succesfully demonstrated in two isolated Michelson interferometers \cite{pradyumna_twin_2020}. Although in this research the use of a single entangled squeezed state did not provide an improvement over independent squeezed states, this result shows the advantages of using various non-classical states of light in measuring correlated signals in interferometers. Therefore, different non-classical states of light can be tested in the co-located 3D interferometers in Cardiff to explore their potential further and enhance the sensitivity to quantum space-time fluctuations.


\section{Detection Statistic}\label{sec:Detection_Statistic} 
As discussed in Sec. \ref{sec:Model_of_Signal}, the sensitivity of the experiment in Cardiff to quantum space-time fluctuations will be achieved through the time integration of the cross-spectral density of the co-located interferometers. 

The primary concern in achieving high sensitivity will be the reduction of noise that is correlated between the interferometers. Unlike uncorrelated noise, correlated noise will lead to an irreducible limit on the experiment's sensitivity. The elimination of such noise comes down to perfecting the isolation between the instruments and removing external noise sources that are coherent on scales greater than the separation between the instruments. 

In the absence of inter-interferometer correlated noise, the detection statistic of the signal will be determined by the magnitude of the signal in the cross spectrum, the uncorrelated noise power spectral densities of both interferometers, the integration time, and the correlation bandwidth of the cross power spectral density. Specifically, the $\chi^2$ detection statistic is given by \cite{gregory_bayesian_2005}
\begin{equation}\label{Eq:chisquared}
    \chi^2= \sum_{i=1}^{n_\mathrm{bin}} \frac{\big(C_{xy}(f_i) - N_{xy}(f_i)\big)^2}{\sigma^2_{N_{xy}}}, 
\end{equation}
 where the null hypothesis is that the cross spectral density $C_{xy}(f)$ contains only known noise with noise cross power spectral density $N_{xy}(f)$ (i.e. ${H_0: \Exp{C_{xy}(f)}=\Exp{N_{xy}(f)}}$). Here, $\sigma^2_{N_{xy}}$ is the variance of $N_{xy}$, the sum is over the frequency bins of the spectrum $f_i$ with ${i\in\{1,2,\,\dots\,,n_\mathrm{bin}\}}$, where the total number of bins $n_\mathrm{bin}$ is determined by the correlation bandwidth $B_{C_{xy}}$ and the length of individual spectra $T_\mathrm{DFT}$. The noise cross power spectral density is the geometric mean of the noise auto power spectral densities in the individual interferometers ($N_{xx}$, $N_{yy}$), and decreases over time with the square root of the number of measured spectra $n_\mathrm{spec}=T_\mathrm{int}/T_\mathrm{DFT}$ \cite{pradyumna_twin_2020,chou_mhz_2017}, i.e.
\begin{equation}
    N_{xy}(f)=\sqrt{\big(N_{xx}(f)\,N_{yy}(f)\big)/n_\mathrm{spec}}.
\end{equation} 
It is expected that the noise spectral densities in the frequency range of interest will be dominated by photon shot noise, which is Gaussian white noise (see Sec. \ref{subsec:shot_noise}), and hence $\sigma^2_{N_{xy}}=\Exp{N_{xy}}^2$. Under the assumption that there is a signal from homogeneous and isotropic quantum space-time fluctuations, and that the separation of the twin interferometers is small compared to the arm length ($d\ll L$), the cross-spectral signal is approximately equal to the auto-spectral signal, $C_{xy}\approx P_{xx}$ (Eq.~\ref{Eq:PSD}). Moreover, it can then be shown that to first order, $\chi^2$ scales as:
\begin{equation}\label{eq:chisquare}
    \chi^2 \appropto  \frac{T_{\rm{int}}\,B_{C_{xy}}\,(L^{\frac{3}{2}-\alpha})^4}{N_{xx}\, N_{yy}}.
\end{equation}    
The experimental design can be understood as an effort to maximise this statistic. An estimate of the time needed to detect or constrain a signal due to quantum space-time fluctuations at a certain confidence level is given in Sec.~\ref{sec:Projected_Sensitivity} by evaluating Eq.~\ref{Eq:chisquared} explicitly using the parameters of the experimental design expounded in Secs.~\ref{sec:Experiment_Design} and \ref{sec:Mitigation_of_Noise}.


\section{Experimental Design}\label{sec:Experiment_Design}
\subsection{Overview}
\begin{figure}[ht]
    \centering
    \includegraphics[scale=0.8]{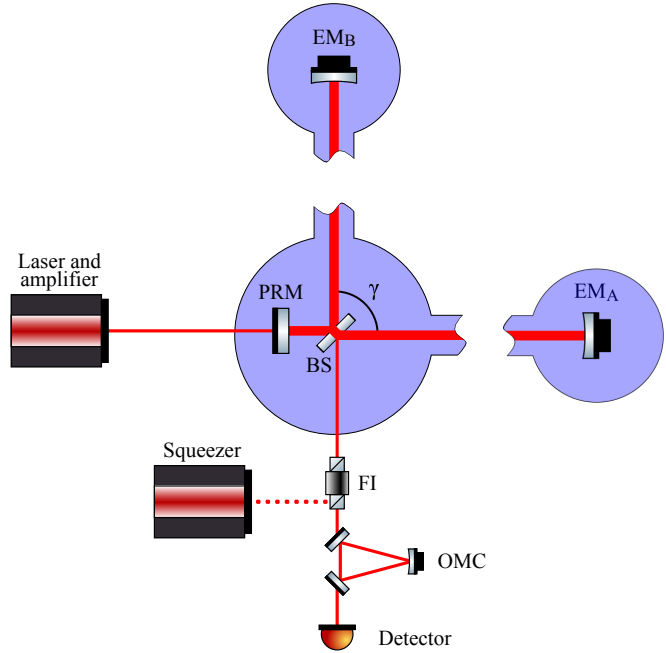}
    \caption{Simplified optical layout of a single power-recycled interferometer. 10 W of input power will be provided by a 1064 nm laser and optical amplifiers, and the laser beam will be matched to the power-recycling cavity using a mode matching telescope. The power-recycling cavity is delimited by the flat power-recycling mirror (PRM) and the concave end mirrors ($\rm{EM}_{\rm{A,B}}$). A 50:50 low-absorption beamsplitter will split and recombine the light in the two arms. The angle between the arms $\gamma$ will depend on the adopted geometry (see Sec. \ref{sec:Model_of_Signal}). At the output of the interferometer, an optical squeezer will be used to inject squeezed vacuum states of light into the interferometer through a Faraday rotator (FI) to reduce quantum noise. An output mode cleaner (OMC) will suppress higher-order optical modes in the output channel. Shaded areas represent vacuum chambers. Most of the arms and their geometry, including the mirrors that link the inner and outer arm segments (see Fig.~\ref{Fig:geometry}), are omitted for simplicity, as indicated by the break symbols.}
    \label{fig:layout}
\end{figure}

An overview of the optical layout of a single interferometer is shown in Fig.~\ref{fig:layout}. The interferometer arms will each have an optical path length of around 3~m. The co-located interferometers are designed to operate with circulating optical powers of roughly 10~kW by using a power-recycling configuration \cite{schnier_power_1997}. 

The input light for each interferometer will be provided by a tunable continuous-wave laser (Coherent Mephisto), with a maximum optical output power of 0.5~W at a wavelength of 1064~nm. A total input power of 10~W will be achieved through the use of optical amplification stages (neoLASE). The optical mode of the laser beam will be matched to the power-recycling cavity of the interferometer with a dedicated mode matching telescope and frequency locked to the common mode interferometer arm length using the Pound-Drever-Hall technique \cite{black_introduction_2000}. The addition of an input mode cleaner to suppress non-resonant optical modes and provide temporal filtering of the input beam is being considered. The power-recycling cavity of each interferometer will be compounded by the 3D interferometer arms, delimited by two concave end mirrors and a single flat power-recycling mirror. Flat mirrors are located along the optical path of each arm and link the inner and outer arm segments. A beamsplitter with one reflective and one anti-reflective surface will split and recombine the light into and from the arms with a 50:50 ratio. 

The interference pattern produced by light from the two arms will be imaged on a photodetector in the output channel. Signals will be obtained using DC-readout \cite{hild_dc-readout_2009,fricke_dc_2012}, and the photodetector outputs will be sampled at 500 MHz using high-frequency digitizers. Between the beamsplitter and the photodetector, an optical mode cleaner (OMC) will be installed to suppress higher-order spatial modes of light in the output channel. In addition, squeezed vacuum states of light, produced using a dedicated setup \cite{vahlbruch_geo600_2010,walls_squeezed_1983}, will be injected into the optical paths of the interferometers through a Faraday rotator, which will allow quantum noise to be reduced (see Sec. \ref{subsec:shot_noise}).

The twin interferometers will each be built in independent but identical vacuum systems that will contain the optics from the power-recycling mirrors to the end mirrors. The rest of the optics, such as the output (and input) mode cleaners, injection optics, and squeezers, will be built in air, to allow for easier manipulation. Each of the twin vacuum systems will consist of five cylindrical vacuum chambers. A single, larger chamber will house the power-recycling-mirror and beamsplitter, and the four slightly smaller chambers will accommodate the mirrors between the arm segments and the end mirrors. To connect the vacuum chambers and allow for alignment, stainless steel vacuum tubes with short bellows at the ends will be used, which connect with adjoining vacuum chambers through manual gate valves. To maximise the options for intervention and adjustment, each vacuum chamber will have its own mechanical (turbo-molecular) pumps, pressure gauges, and valves for vacuum isolation. This feature will contribute to the fast reconfigurability of the 3D geometry of the interferometers. 

\subsection{Improvements over Previous Experiments}
The use of co-located interferometers to search for signatures of quantum gravity is not new; the same concept has been employed in the Fermilab Holometer. The Holometer programme, which started in 2012, uses two co-located Michelson interferometers with a 40~m arm length to search for signatures of quantum space-time \cite{chou_holometer_2017}. The individual interferometers of the Holometer have reached a sensitivity to displacements of the order of $10^{-18}\,\mathrm{m}/\sqrt{\rm Hz}$ at frequencies between 1 and 25~MHz, limited by photon shot noise. The cross-spectrum of the two interferometers was averaged over time and examined for the presence of correlated noise due to quantum space-time fluctuations. The experiment did not detect any unexplained correlated signals, and thus constrained the possible magnitude of longitudinal quantum space-time fluctuations \cite{chou_interferometric_2017}\footnote{In its original configuration, the Holometer used a classical Michelson arm geometry (corresponding to $\gamma=90^\circ$, $\beta=180^\circ$, $\omega=0^\circ$ in Fig.~\ref{Fig:geometry}). Therefore, this result specifically set an upper limit on potential longitudinal quantum space-time fluctuations that are anti-correlated at an angle of $90^\circ$ and fully coherent between identical instruments separated by $\approx$~1~m. This is because the Holometers's interferometer arms are orthogonal, and a signal would only have been detected in the cross-spectrum
for length changes that are anti-correlated between the two arms of each interferometer, and coherent between the two interferometers.}.

The main improvements to the design that will allow the experiment in Cardiff to surpass the displacement sensitivity of the Holometer are the inclusion of optical mode cleaners, higher optical input power, and the injection of squeezed states of light, which all will allow for the mitigation of photon shot noise (see Sec. \ref{subsec:shot_noise}). 

The smaller scale interferometry experiment at Cardiff University will have a reconfigurable arm geometry as considered in Sec.~\ref{sec:Model_of_Signal}, with interferometer arms that include a radial and a non-radial segment to introduce a sensitivity to transverse quantum space-time fluctuations. To resolve the signal described by Eq. \ref{Eq:PSD}, the output needs to be sampled with a frequency higher than the inverse light crossing time $c/L$, such that the signal of single photon traversals of the interferometer arms is not averaged with that of subsequent ones \cite{kwon_interferometric_2016}. Therefore, a high-frequency data acquisition and analysis system will be built, both to sample the photodetector output at up to 500 MHz, and to Fourier transform and correlate the data.

The shorter optical path length compared to the Fermilab Holometer, which would theoretically produce a weaker signal, is compensated by an increased correlation bandwidth of 250 MHz compared to the Holometer's 25 MHz. As can be seen from Eq.~\ref{eq:chisquare}, this is possible because the numerator of the detection statistic depends on the product of the optical path length and the bandwidth.  

The compact table-top design allows the experiment to be built on a single vibration-isolated optical bench in a laboratory under stable environmental conditions. Moreover, the small scale of the instruments allows, at least in principle, for a more rapid reconfiguration of the geometry. This makes it possible to perform measurements using different arm geometries (see Fig. \ref{Fig:geometry}), which provides the possibility to probe the angular correlations of quantum space-time fluctuations. 

As detailed below, the noise spectral densities $\rm{NSD}_{1,2}$ will be reduced below the levels observed in the Fermilab Holometer, and it is thus expected that the experiment in Cardiff will yield a sensitivity to longitudinal and transverse quantum space-time fluctuations that exceeds that of the Fermilab Holometer. For comparison, the main experimental parameters for both the Holometer and the twin interferometers to be built in Cardiff are summarised in Table~\ref{tab:sec5_summary}.

\begin{table}[h]
\centering \small
\renewcommand{\arraystretch}{1.2}
\begin{tabular}{|l||c|c|}
\hline 
\multicolumn{1}{|c||}{\textit{Parameter}} & \multicolumn{1}{c|}{\textit{Fermilab}} & \multicolumn{1}{c|}{\textit{Cardiff (planned)}}      \\ \hhline{|=||=|=|}
Optical path length & 40 m & 3 m \\ \hline
Input power & 2 W  & 10 W\\ \hline
Circulating power & 2 kW & 10 kW\\ \hline
Bandwidth& 25 MHz & 250 MHz \\ \hline
Effective CD     & $2\cdot10^{-5}$ & $<10^{-6}$ \\ \hline
Detected SQZ (initial/final target) & --  & 6/10 dB   \\ \hline
\hline
\end{tabular}
\caption{The main experimental parameters for both the Fermilab Holometer and the twin interferometers to be built in Cardiff are summarised. Here, `Effective CD' is the effective contrast defect, defined as the ratio of the power as measured on the photodetector and the circulating power at the beamsplitter on a minimum of the interference pattern. This is different from the more conventional definition of the CD in Sec.~\ref{sec:Mitigation_of_Noise}, which would not allow for a direct comparison of the CD due to the absence of an OMC in the Fermilab Holometer. Detected SQZ is the level of light squeezing as detected at the output; squeezing is not employed in the Fermilab Holometer.} 
\label{tab:sec5_summary}
\end{table}


\section{Mitigation of Noise}\label{sec:Mitigation_of_Noise}
Decreasing the level of uncorrelated stochastic noise, which amounts to minimising the denominator in Eq. \ref{eq:chisquare}, is a primary strategy towards increasing the sensitivity. In this section, the sources of different kinds of noise and the strategies for their mitigation are discussed, as summarised in Table~\ref{tab:noise_mitigation}.

\begin{table}[h]
\centering \small
\renewcommand{\arraystretch}{1.2}
\begin{tabular}{|l||l|}
\hline 
\multicolumn{1}{|c||}{\textit{Noise Source}} & \multicolumn{1}{c|}{\textit{Mitigation Strategy}}                                                        \\ \hhline{|=||=|}
Shot noise        & \begin{tabular}[c]{@{}l@{}}High input power, power recycling, \\ squeezed states of light\end{tabular}   \\ \hline
Additional shot noise due to contrast defect           & \begin{tabular}[c]{@{}l@{}}High-performance optics and alignment,\\ output mode cleaner\end{tabular}     \\ \hline
Seismic and acoustic noise                  & \begin{tabular}[c]{@{}l@{}}Vibration-isolated optical bench, \\ quiet cleanroom environment\end{tabular} \\ \hline
 Residual gas noise                          & Ultra-high-vacuum system                                                                                \\ \hline
Noise from stray light                      & Baffles in vacuum tubes (optional)                                                                                 \\ \hline
Thermal noise                               & No mitigation needed                                                                                           \\ \hline
\end{tabular}
\caption{Relevant sources of noise and strategies for their mitigation.}
\label{tab:noise_mitigation}
\end{table}

\subsection{Shot Noise}\label{subsec:shot_noise}
It is evident in similar state-of-the-art interferometry experiments, specifically gravitational wave interferometers and the Fermilab Holometer, that the maximum experimentally attainable sensitivity in the MHz range is currently limited by photon shot noise.

The origin of this type of noise is the Heisenberg uncertainty relation for the particle number and phase observables of a photon state
\begin{equation}
    \Delta N_{\gamma}\, \Delta  \phi \geq1,
\end{equation}
 where $\Delta N_{\gamma}$ and $\Delta \phi$ are the root mean square (RMS) uncertainty of the number of photons $N_\gamma$ and their phase $\phi$ with respect to a coherent state. The RMS uncertainty of the number of photons is described by Poisson statistics as $ \Delta N_{\gamma}=\sqrt{N_{\gamma}}\equiv\sqrt{(P\tau)/(\hbar \omega)}$, where $P$ is the expectation value of the optical power of the measured state, $\tau$ is the total measurement time, and $\omega$ the angular frequency of the light. The uncertainty of the measured phase is $\Delta \phi \equiv(2\pi/\lambda)\Delta L $. Thus, there is a lower limit on the uncertainty of measured length $\Delta L \geq \lambda /(2\pi \Delta N_{\gamma})$. This leads to a displacement noise amplitude spectral density in the output of \cite{grote_advanced_2019,caves_quantum-mechanical_1981}   
\begin{equation}\label{eq:SNPSD}
S^{\text{SN}}_{\Delta L}(f)= \sqrt{ \frac{ c \hbar\lambda}{4 \pi P_{BS}}},
\end{equation}
which is constant over all frequencies, and can be seen to depend inversely on the circulating power at the beamsplitter $P_{\rm{BS}}$. 

Power recycling will be employed in the interferometers to reach high circulating optical powers, which allows the effective mitigation of shot noise \cite{regehr_demonstration_1995}. The power-recycling cavities will be linear plano-concave, with a common flat Power Recycling Mirror (PRM) as the input optic and end mirrors with a radius of curvature (RoC) of 6 m.

The steady-state circulating optical power at the beamsplitter of a power-recycled interferometer as in Fig.~\ref{fig:layout} is
\begin{equation}\label{eq:powerrecycling}
P_{\rm BS} = \frac{4\, P_{\rm IN}\, T_{\rm PR}}{\left[2-(\sqrt{1-T_{\text{PR}}- A_{\text{PR}}}) \left(\sqrt{R_{\rm{EM_A}}}R_{\rm{MM_A}}+\sqrt{R_{\rm{EM_B}}}R_{\rm{MM_B}}\right) \left(1- A_{\text{BS}}\right) \right]^2},
\end{equation}
where $P_{\rm IN}$ is the input power at the PRM, $A_{\rm PR}$ and $T_{\rm PR}$ are respectively the absorption and transmittance of the PRM, $A_{\rm BS}$ parameterises the absorption and losses from the beam splitter, $R_{\rm{MM_{A,B}}}$ are the reflectances of the mirrors between the inner and outer segment, and $R_{\rm{EM_{A,B}}}$ are the reflectances of the end mirrors of the arms. Losses due to scattered light are accounted for in $A_{\rm{BS}}$ for simplicity. 

As is clear from Eq.~\ref{eq:powerrecycling}, the power at the beamsplitter depends linearly on the incident power, which motivates the use of injected optical power of up to 10~W. Limiting factors in achieving a high circulating power are absorption in the optics and the generation of higher-order optical modes (HOMs \footnote{The fundamental Gaussian optical mode (TEM00) is the lowest order solution of the paraxial wave equation. An infinite number of higher-order solutions exist which are referred to as higher-order modes \cite{siegman_mode_1974}.}), which will be discussed below. To attain the desired circulating power on the beamsplitter of $P_{\rm BS}=10$~kW with an input power of $P_{\rm IN}=10$~W, $R_{\rm EM_{A,B}}$, $R_{\rm MM_{A,B}}$ will need to be $\geq 0.9997$, assuming equal reflectance for all arm mirrors. This value applies for a design with an ultra-low absorption ($\leq0.5\, \rm{ppm}/\rm{cm}$) beamsplitter, and using a PRM for which $A_{\rm PR}=10^{-5}$, and $T_{\rm PR}=5\cdot10^{-4}$.

In order to improve the detector sensitivity beyond the shot noise level of Eq.~\ref{eq:SNPSD}, squeezed states of light will be used to mitigate the shot noise by decreasing the uncertainty in the phase $\Delta\phi$ at the detector \cite{walls_squeezed_1983}. Independent or entangled squeezed vacuum states will be injected into the anti-symmetric ports of the interferometers \cite{pradyumna_twin_2020}. The initial target is to reach a level of detected squeezing of around 6~dB (this level has recently been reached in GEO\,600 \cite{lough_first_2020}, and is double that currently used in Advanced LIGO \cite{tse_et_al_quantum-enhanced_2019}). A goal for a future stage will be to achieve 10~dB of detected squeezing.

\subsubsection{Additional Shot Noise due to Contrast Defect}\label{subsubsec:CD_and_HOMs}
In a realistic interferometer operated at the dark fringe, the expectation value of the measured power at the anti-symmetric port is not zero due to differential imperfections in the arms.  These conditions lead to additional shot noise on top of the level given by Eq.~\ref{eq:SNPSD}. Therefore, the signal-to-shot-noise ratio (SNR) decreases with this non-signal-carrying power at the anti-symmetric port $P_\text{AS}$ as
\begin{equation}
    \mathrm{SNR} \propto \frac{P_\text{sig}}{\sqrt{P_\text{BS}+P_\text{AS}+P_\text{sig}}} = \frac{a_\text{sig}\sqrt{P_\text{BS}}}{\sqrt{1+\text{CD} +a_\text{sig}}}
\end{equation}
where $P_{\text{sig}}=a_{\text{sig}}P_{\text{BS}}$ is the signal power at the anti-symmetric port, $a_{\text{sig}}$ is the dimensionless signal amplitude, and $\text{CD}$ is the contrast defect, which is defined here as $\mathrm{CD}\equiv P_\text{AS}/P_\text{BS}$ as measured at a minimum of the interference pattern. The CD can be expanded in its dominant contributions as
\begin{equation}
    \mathrm{CD} \approx c_{\Delta R} (\Delta R)^2 + c_{\Delta A} (\Delta A)^2 + c_{\Delta\alpha}\Delta\alpha + G_{\text{HOM}} + \dots 
\end{equation}
Here, $\Delta R$ and $\Delta A$ are the differences in reflectance and absorption between the two arms, respectively, $\Delta\alpha$ is the difference in birefringence between the two arms, $c_x$ are numerical coefficients of order 1, and $G_{\text{HOM}}$ parameterises the CD due to HOMs. The former terms are due to imperfect destructive interference of light in the fundamental Gaussian mode (TEM00), whereas $G_{\text{HOM}}$ is due to imperfect destructive interference of HOM light. The experiment will be designed with a target of attaining a CD due to TEM00 light~$<10^{-6}$, and a CD due to HOM light~$<10^{-5}$. The latter requirement is less stringent, as the contribution of HOM light will be reduced by at least an order of magnitude at the output using an optical mode cleaner (as detailed below). As a reference, the lowest CD reported in the Fermilab Holometer was about $2\cdot10^{-5}$ \cite{chou_holometer_2017}. Computer simulations using FINESSE \cite{freise_finesse_2013} have been carried out to quantify the expected CD due to differential optical properties of the arms, some of which result in the generation of HOMs.

Reflectance differences between the arms were simulated, showing that the optical components that will be used have to be within a reflectance tolerance of $0.2~\%$ to achieve a CD due to TEM00 light~$<10^{-6}$. Differential absorption losses in the coatings of the end mirrors, as well as losses on the anti-reflective surface of the BS (light from only one of the arms interacts with this surface) will create a negligible contribution to the overall CD. Simulations demonstrate that as long as losses on any one surface are kept to below 100 ppm, the resulting CD will be $\sim10^{-9}$.

Inter-arm differences in the radius of curvature of the end mirrors generate a mode mismatch between the arms of the interferometer, resulting in the transfer of power from the fundamental mode to second order HOMs at the output. It was inferred from simulations that discrepancies in RoC should be limited to $0.3~\%$ to attain a CD due to HOM light~$<10^{-5}$. 

Slowly deteriorating misalignment of the interferometer mirrors due to experimental conditions, called drift, is another source of HOMs at the anti-symmetric port. A FINESSE simulation found that misalignment of all arm mirrors in opposite directions, which is the worst-case scenario, by as little as $0.2\,\mathrm{\mu}\rm{rad}$ each results in a contrast defect exceeding $10^{-5}$ (misalignments of single arm mirrors or misalignments of multiple arm mirrors in the same direction produce a much smaller CD). Therefore, to achieve the target CD due to HOMs of $<10^{-5}$, the arms have to be kept in alignment to well within $0.2\,\mu\rm{rad}$. To realise this, it is expected that a feedback control system is required. 

Thermal lensing \cite{winkler_geo_2007} also produces a CD through HOM generation, with the dominant effect expected from the beamsplitter. Due to power absorption within the beamsplitter there is a temperature gradient generated within the medium that generates a corresponding gradient in the refractive index, which causes lensing. Since the effective thermal lens is proportional to the absorption within the substrate \cite{winkler_heating_1991}, it is desirable to select a relatively thin beamsplitter to minimise the power absorption. For the beam sizes planned in this experiment (which are small compared to those in e.g. gravitational wave detectors) a very thin beamsplitter can be used. For fused silica with ultra-low absorption (0.5~ppm/cm) and a beamsplitter thickness of 6~mm, the HOM power at the anti-symmetric port due to thermal lensing is calculated to be only 0.2~mW, which corresponds to a CD of $6\cdot 10^{-8}$\footnote{For reference, if a  9~cm thick beamsplitter (such as the one employed at GEO\,600 \cite{winkler_geo_2007}) is considered in the same setup, it is estimated that roughly 1~W of HOM light will be present at the anti-symmetric port, which would require the addition of compensation plates in the experiment.}.

Differential birefringence entails a difference of the optical path length of the two orthogonal polarisation states that is not balanced between the two arms. As the photo-detector is insensitive to the polarisation of light at the output, the detected interference pattern results from a superposition of interference patterns of both polarisation states and will therefore have imperfect contrast. If necessary, the differential birefringence could be minimised by rotating the birefringence axes along the direction of the light polarisation \cite{ejlli_pvlas_2020}. In principle, the variable birefringence as produced in the Cotton-Mouton effect or the Faraday effect could be used to calibrate the apparatus.       

\subsubsection{Suppression of HOM Light at the output using an OMC}
The measures described above should ensure a consistent reduction of the HOM power. However, the finite manufacturing tolerances of the optics, their degradation over time, and thermal effects, will still lead to non-negligible HOM generation.

HOM light at the anti-symmetric port is a significant part of the CD in state-of-the-art interferometry experiments, such as gravitational wave detectors
and the Fermilab Holometer. These interferometers use DC-readout, and as such operate at a small offset from the dark fringe \cite{hild_dc-readout_2009,fricke_dc_2012}. Ideally, the magnitude of the offset
is chosen such that the CD represents an insignificant fraction of the power incident on the photodetector compared to the TEM00 light due to this intentional dark fringe offset. If there is a significant amount of HOM light at the output, this would require operating at a large offset, which has the disadvantage that noise from other sources is more strongly coupled to the output \cite{hild_dc-readout_2009,fricke_dc_2012}. In addition, a large offset to dominate significant amounts of HOM light would entail high power incident on the photodetector, which is undesirable, as high-bandwidth photodetectors are subject to technical limitations at high incident power. This effectively constrains the circulating power that can be used in the interferometers, which in turn limits the shot-noise mitigation that can be achieved. 

To address these issues, an output mode cleaner (OMC) will be included in the output channel of each interferometer (see Fig.~\ref{fig:layout}), with the target of suppressing HOM power by at least an order of magnitude (OMCs are also employed in gravitational wave detectors, but not in the Fermilab Holometer). The addition of the OMCs will reduce the amount of HOM light incident on the photodetectors, thereby facilitating the use of a smaller dark fringe offset. The OMCs will thus not only improve the shot-noise-limited sensitivity directly by suppressing the HOM power, but also by allowing higher circulating power within the interferometers.

Preliminary modelling indicates a cavity finesse ($\mathcal{F}$) of $>\,$11 is required to achieve a suppression of the HOM power at the output of an order of magnitude. A large cavity bandwidth ($B_c$) of $\approx 200\; \rm{MHz}$ is required to maintain a high output signal bandwidth (see Sec. \ref{sec:Detection_Statistic}). This large bandwidth coupled with the finesse requirement leads to an upper limit on the cavity round trip length $l_{rt}$ of around 13.5~cm, as $l_{rt}= c/(\mathcal{F}B_{c})$. Stronger suppression of HOM light can be achieved, but requires a higher finesse and thus a shorter round trip length (for example, a finesse of 30 gives suppression by a factor~$>\,$60 with a $l_{rt}$ of $\approx\,$5~cm). As such, a triangular cavity geometry is preferable to a bow-tie alternative, to keep physical size as large as possible while maintaining a short optical path. The relevant parameters for the suppression of HOM light using an OMC are summarised in Table~\ref{tab:sec6_summary}.

\begin{table}[h]
\centering \small
\renewcommand{\arraystretch}{1.2}
\begin{tabular}{|l||c|}
\hline 
\multicolumn{1}{|c||}{\textit{Parameter}} & \multicolumn{1}{c|}{\textit{Value}}  \\
\hhline{|=|=|}
CD target (TEM00/HOM light)       & $<$10$^{-6}$/$<$10$^{-5}$   \\ \hline  
OMC HOM power suppression factor       & $\geq10$   \\ \hline
OMC Finesse    & $>$ 11   \\ \hline
OMC bandwidth    & 200 MHz   \\ \hline
OMC round-trip length           & $<$ 13.5 cm  \\ \hline
\end{tabular}
\caption{Experimental parameters concerning the suppression of shot noise due to contrast defect from higher order optical modes using an optical mode cleaner. The CD is the ratio of the circulating power and the power at the ant-symmetric port as measured on a dark fringe, before the OMC. The OMC will suppress HOM light by at least an order of magnitude to yield an effective CD at the output of $<10^{-6}$ such that the experiment will be limited by shot noise due to vacuum fluctuations.}
\label{tab:sec6_summary}
\end{table}

\subsection{Seismic and Acoustic Noise}
Seismic vibrations, air turbulence, and acoustic noise cause the most problematic low-frequency noise in interferometry, as these sources produce phase noise through various coupling mechanisms. The RMS displacement due to environmental noise of the vibration-isolated optical bench on which the experiment will be built was measured, and is less than $10^{-11} \,\rm{m/\sqrt{Hz}}$ at 100 Hz.

Since seismic noise is filtered at higher frequencies by the vibration-isolated optical bench, the effect of seismic noise on the displacement amplitude spectral density of the interferometer at frequencies above 1~MHz is expected to be below that of the shot noise by more than an order of magnitude. This extrapolation, supported by reported seismic noise levels at high frequencies in other interferometry experiments, suggests that no further mitigation of seismic noise will be required in the experiment. Environmental noise on the input beam due to air turbulence and dust will be mitigated by placing the whole apparatus in a quiet Class 10000 cleanroom with an air circulation and filtration system.

\subsection{Noise from Residual Gas}
The jitter in the column density of the residual gas in the vacuum system causes fluctuations of the absolute phase of the light which will manifest as phase noise in the fundamental optical output mode. To quantify this issue the approach taken by the LIGO Collaboration was followed \cite{zucker_proceedings_1996}. In that work it was found that the magnitude of optical path length fluctuations due to residual gas molecules passing through the beam scales with the square root of the pressure. Moreover, it was found that this type of noise has a roughly flat frequency spectrum for frequencies up to the inverse duration of the average transition of a gas molecule through the beam diameter ($\approx 0.1$~MHz for the planned beam size and ambient temperature), and drops off exponentially for higher frequencies. The residual gas displacement noise as estimated using the results in \cite{zucker_proceedings_1996} for a vacuum pressure $<10^{-6}$~mbar will be less than the displacement shot noise by more than one order of magnitude at all frequencies. 

Another mechanism through which residual gas degrades the sensitivity is the absorption of light by hydrocarbons deposited onto optical surfaces. The deposition rate depends linearly on the residual gas pressure, as detailed in \cite{chou_holometer_2017}.
To increase the time needed to deposit a single hydrocarbon monolayer on an optical surface to six months, which corresponds to an increment of the coating absorption of 20~ppm, a vacuum with a pressure of~$\leq 10^{-7}$~mbar will be used \cite{chou_holometer_2017}. Furthermore, the contamination of the internal surfaces of the vacuum system will be kept sufficiently low before baking.

\subsection{Noise from Stray Light}
Another non-negligible source of noise is stray light produced by specular and diffuse reflections of light by dust or imperfections on the optical components. The noise due to stray light has been thoroughly studied for the LIGO interferometers. In particular, it was found that the coupling of seismic noise to vibrational excitations of interferometer components leading to phase modulation of stray light was a significant source of noise \cite{e_noise_1994}. 
The table-top interferometers in Cardiff will be sensitive at much higher frequencies than ground-based gravitational wave detectors such as LIGO. Therefore, we expect the contribution of noise due to stray light modulated by seismic noise (which falls off at higher frequencies, see above) to be negligible.

However, stray light may still reach the detector through multiple grazing reflections from the internal surface of the vacuum chambers, and thus produce a contrast defect. In the LIGO interferometers, the installation of baffles inside the vacuum tubes was used to attenuate stray light. This solution works by blocking the scattered path rays reaching the internal surface of the tube at a grazing angle. 

The baffles are positioned in such a way that stray light produced on the mirrors will never reach the internal surface of the vacuum tube. Specifically, the position of the $n^{\rm{th}}$ baffle $l_n$ is given by the geometric series $l_n=l_0\beta ^n$, where $\beta >1$ is given by the radius of the vacuum tube relative to the height of the baffles \cite{e_noise_1994}.

The installation of baffles in this arrangement to suppress stray light is being considered for the experiment in Cardiff. In addition, the baffles may also attenuate back-scattered light in very high order optical modes from the OMC, which would contribute to the overall suppression of HOM light in the interferometer. 

\subsection{Thermal Noise}
The thermal excitation of atoms, or Brownian motion, excites the mechanical vibration modes of optical elements and as a consequence generates displacement fluctuations of the reflective surfaces. Given the dimensions and material properties of the optical elements, it is expected that the fundamental vibration mode and its harmonics lie at $\approx$~477~kHz and multiples thereof.
As the thermal noise spectrum is expected to consist of a number of narrow resonance peaks, it need not be mitigated. These peaks can be excluded in the data analysis with a small reduction in useful signal bandwidth.


\section{Projected Sensitivity}\label{sec:Projected_Sensitivity}
\begin{figure}[ht]
\centering
\includegraphics[scale=0.45]{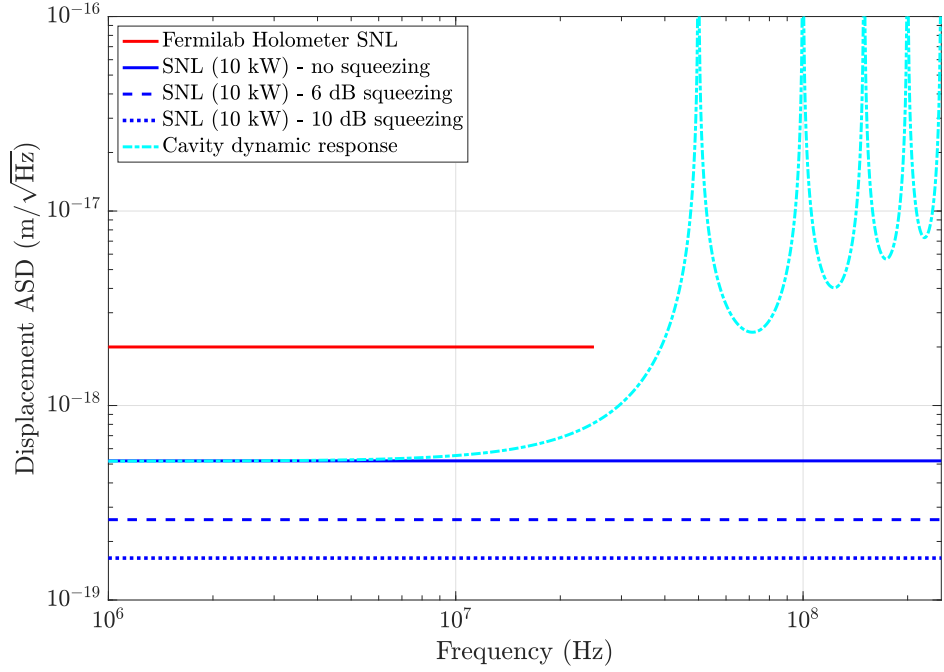}
\caption{Comparision of the shot-noise-limited displacement noise amplitude spectral densities (SNL) of a single Fermilab Holometer interferometer and the projected SNL of a single Cardiff interferometer using different levels of squeezing. Curves plotted are the reported final single interferometer SNL of the Fermilab Holometer (solid red line), projected single interferometer SNL for the experiment in Cardiff using 10~kW of circulating power with no squeezing (solid blue line), 6~dB of squeezing (dashed blue line), and 10~dB of squeezing (dotted blue line). In addition, the projected inverse sensitivity to harmonic cavity length changes without squeezing is shown (dash-dotted cyan curve).}
\label{Fig:displacement_sqz}
\end{figure}

The sensitivity of the experiment in Cardiff is designed to be limited by shot noise due to vacuum fluctuations. All other sources of noise, except for thermal noise peaks from the optics, will be mitigated to below this shot noise floor as detailed in Sec.~\ref{sec:Mitigation_of_Noise}. Based on Eq.~\ref{eq:SNPSD} and using the target circulating power $P_{\rm{BS}}=10\;\rm{kW}$, the estimated shot-noise-limited displacement amplitude spectral density (or shot noise level, SNL for short) for a single interferometer is $5.17\cdot10^{-19} \,\rm{m}/\sqrt{\rm{Hz}}$ without squeezing. The estimated shot-noise-limited displacement noise amplitude spectral density is plotted in Fig.~\ref{Fig:displacement_sqz} for different levels of squeezing. 

In addition, a curve generated using FINESSE \cite{freise_finesse_2013} is plotted in Fig.~\ref{Fig:displacement_sqz} that shows the inverse shot-noise-limited sensitivity to phenomena that harmonically modulate the distance between the cavity mirrors, such as gravitational waves. It can be seen that the sensitivity decreases at every multiple of the free spectral range of the cavity. This is because the measured length change will be zero if an oscillatory signal changes phase by $k\pi$ ($k\in \mathbb{Z}$) within a single light crossing time $L/c$. 

By taking the SNL to equal the total expected noise amplitude spectral density $\sqrt{N_{xy}}$, Eq~\ref{Eq:chisquared} can be evaluated explicitly for the experimental design, assuming a specific scaling and magnitude of quantum space-time fluctuations. An estimate for the integration time required to detect or constrain a quantum gravity signal of the form of Eq.~\ref{Eq:PSD} at a certain confidence level can then be made.

For the current estimate, the quantum space-time fluctuations are assumed to behave as homogeneous and isotropic white noise with a variance equal to the Planck length (i.e. $\alpha=1/2$, $A_{ij}=\delta_{ij}$). The power spectral density of the signal is then $P_{xy}\sim10^{-43}\,\mathrm{m}^2/\rm{Hz}$ (Eq~\ref{Eq:PSD}), which is further assumed to be fully coherent between the twin interferometers, i.e. $C_{xy}=P_{xy}$. Note that such a signal is currently constrained to roughly this magnitude for an inter-arm angle of $\gamma=90^\circ$ by the results of the Fermilab Holometer \cite{chou_interferometric_2017}), and that constraint thus only applies under the additional assumption that the fluctuations are anti-correlated at $90^\circ$ (see Sec.~\ref{sec:Experiment_Design}).

The experimental parameters factor in to the estimate through the noise level and correlation bandwidth. The correlation bandwidth is determined by the output sampling frequency, which is 500~MHz for the current design. The noise auto power spectral densities in both interferometers are assumed to be equal to the shot noise power spectral density for 10~kW of circulating power and 6~dB of squeezing, thus $\Exp{N_{xx}} =\Exp{N_{yy}}=\Exp{N_{xy}}=(S_{\Delta L}^\mathrm{SN})^2\approx 7\cdot10^{-38}\,\mathrm{m}^2/\rm{Hz}$. 

Consequently, the estimated integration time to detect or constrain such a quantum gravity signal at a confidence level of $5\sigma$ is $\approx5\cdot10^5$~s. For a duty cycle of $80\%$, this would require roughly a week of operating the experiment.


\section{Summary}\label{sec:Conclusions}
Holographic theories of quantum gravity predict that measurements of distance will exhibit fluctuations with macroscopic correlations. This phenomenological hypothesis has been translated into a quantitative prediction for the power spectral density of a signal due to these fluctuations in a 3D interferometer, as a function of the scaling and anisotropy of the fluctuations and the geometry of the arms.  

Co-located table-top 3D interferometers will be built at Cardiff University to detect or set upper limits on such signatures of quantum gravity. Time integration of the cross-spectrum of the twin instruments is expected to allow for increasing signal-to-noise ratios, as the quantum geometrical fluctuations are expected to be correlated between the instruments, whereas dominant sources of noise are not.  

In addition, this experiment will be sensitive to high-frequency gravitational waves and various dark matter candidates. 

To reach a displacement sensitivity limited by photon shot noise, environmental and internal noise will be strongly mitigated. High power lasers and power recycling will be employed to reach circulating optical powers of 10~kW, and an optical output mode cleaner will be used to suppress photon shot noise from higher order optical modes. To go beyond the shot noise level, independent or entangled squeezed states of light will be used in the interferometers. 

The projected sensitivity of the experiment was estimated using simulations and models validated by observations in other precision interferometers, such as gravitational wave detectors. The interferometers being built will be sensitive to displacements on the order of $10^{-19}\,\rm{m}/\sqrt{\rm{Hz}}$ in a frequency band between around 1 and 250~MHz, thus surpassing the sensitivity of previous interferometry experiments in this frequency band. Using this displacement sensitivity, it was estimated that the integration time required for the experiment to detect or constrain a signal from isotropic and homogeneous quantum-space time fluctuations with a variance on the order of the Planck length at a confidence level of $5\sigma$ is $\approx5\cdot10^5$~s.


\section*{Acknowledgments}
We thank the Leverhulme Trust for support of this work under grant RPG-2019-022, and we thank the Science and Technology  Facilities Council (STFC) for support of this work. The authors thank Ohkyung Kwon and Craig Hogan for suggestions and discussions on the design of the experiment. S.M.V. would like to thank Kathryn Zurek for discussion on the theory of quantum space-time fluctuations. L.A. would like to thank Anna Green for her support and availability in fixing some minor FINESSE bugs.

\section*{References}

\bibliography{references.bib}
\bibliographystyle{custombstv4.bst}

\end{document}